\documentclass{article}
\usepackage[preprint, nonatbib]{neurips_2024}

\usepackage[makeroom]{cancel}
\usepackage[numbers]{natbib}
\usepackage{hyperref}
\usepackage{amsmath}
\usepackage{amssymb}
\usepackage{enumitem}
\usepackage{tikz}
\usepackage{booktabs}
\usepackage{amsfonts}
\usepackage{dsfont}
\usepackage{subcaption}
\usepackage{tabularx}

\title{Kirigami: large convolutional kernels improve deep learning-based RNA secondary structure prediction}

\author{
    Marc Harary \\
	Department of Data Science \\
	Dana-Farber Cancer Institute \\
	Boston, MA 02115, USA \\
	\texttt{marc@ds.dfci.harvard.edu}
	\And
    Chengxin Zhang \\
	Department of Computational Medicine and Bioinformatics \\
	University of Michigan \\
	Ann Arbor, MI 48109, USA \\
	\texttt{zcx@umich.edu}
}

\begin{document}

\maketitle

\begin{abstract}
	We introduce a novel fully convolutional neural network (FCN) architecture for predicting the secondary structure of ribonucleic acid (RNA) molecules. Interpreting RNA structures as weighted graphs, we employ deep learning to estimate the probability of base pairing between nucleotide residues. Unique to our model are its massive 11-pixel kernels, which we argue provide a distinct advantage for FCNs on the specialized domain of RNA secondary structures. On a widely adopted, standardized test set comprised of 1,305 molecules, the accuracy of our method exceeds that of current state-of-the-art (SOTA) secondary structure prediction software, achieving a Matthews Correlation Coefficient (MCC) over 11-40\% higher than that of other leading methods on overall structures and 58-400\% higher on pseudoknots specifically.
\end{abstract}

\section{Introduction}
In the context of ribonucleic acids (RNAs), secondary structure prediction refers to the inference of the set of nucleotide base pairs corresponding to a given primary structure \cite{1}. Close parallels exist to contact prediction in protein structure modeling, where deep learning methods have been successful \cite{2,3}. On the other hand, the comparative scarcity of high-quality data poses a significant challenge for the computational analysis of RNA secondary structure \cite{4}, which has prompted the development of a wide gamut of algorithms that draw from evolutionary biology, thermochemistry, statistics, and deep learning to circumvent this limitation \cite{4, 5, 6, 7, 8, 9, 10, 11}.

Covariance techniques like R-scape \cite{5} and PLMC \cite{6} exploit comparative phylogenetic analysis to identify pairwise co-occurrence of structural motifs across multiple sequence alignments, though the shortage of available data limits the reliability with which evolutionarily conserved structures can be identified. Thermodynamic methods like RNAfold from ViennaRNA \cite{7}, RNAstructure \cite{8}, ContextFold \cite{12}, and mfold \cite{9} instead assign annotations that minimize free energy (MFE) estimated from parameters in the Nearest Neighbors database (NNDB) \cite{13}. Their primary weakness lies in assuming a strict equivalence between the empirically likeliest structure and the most energetically stable, given that a non-trivial minority of molecules (e.g., snoRNAs) exist that do not occupy MFE states \cite{14}. Moreover, thermochemical algorithms typically employ dynamic programming \cite{15}, which cannot predict non-nested base pairs and thereby diminishes accuracy on pseudoknotted molecules \cite{16} (Figure 2G-H). An additional class of traditional RNA algorithms, stochastic context-free grammars like those of PPfold \cite{10}, borrow from computational linguistics to model structural motifs like hairpin loops and bulges as syntactical elements of a formal language, although they are also highly dependent on empirical data for statistical estimates \cite{14} and limited to nested structures \cite{16}. Finally, statistical models like IPknot \cite{17} employ other means of scoring secondary structures based on empirical likelihood to estimate the maximum expected accuracy (MEA) solution, with some methods like PETfold \cite{14} unifying multiple approaches.

Recently, several pipelines based on deep learning have obtained results in many cases superior to those of the foregoing \cite{4,11,18}. SPOT-RNA \cite{4} employs transfer learning \cite{19} and an ensemble of models composed of long short-term memory \cite{20,21}, residual neural network (ResNet) \cite{22}, and multilayer perceptron \cite{23} layers to achieve relatively high accuracy on both nested and non-nested structures. Interpolating between biophysical and neural algorithms, MXfold2 \cite{11} consists of a smaller, though architecturally similar model with convolutional \cite{24} layers trained via a sophisticated loss function that accounts for the NNDB parameters. While its performance on nested structures is accordingly high, its reliance on dynamic post-processing limits performance on pseudoknots. E2Efold’s \cite{25} end-to-end differentiable algorithm obtains poor results in comparison to approaches based either in deep learning or otherwise \cite{11,18}. Most recently, UFold \cite{18} achieves competitive accuracy by means of convolutional and deconvolutional layers inspired by UNet \cite{26}, a model originally designed for segmenting medical images.

Departing from the former architectures, the following work introduces Kirigami, a fully convolutional network (FCN) \cite{27} trained on secondary structure annotations and architecturally inspired by recent developments in computer vision \cite{28}. It has been demonstrated that larger kernel sizes and an absence of pooling and deconvolutional layers yield state-of-the-art (SOTA) results for image classification \cite{28}, at least in some cases even outperforming attention-based models like vision transformers \cite{29}. By adapting these innovations and expanding kernel sizes even further beyond those of most existing convolutional networks for computer vision, we create a powerful model for RNA secondary structure prediction that exceeds the performance of several SOTA algorithms on a well-established test set \cite{4,11,18}. Kirigami is an open-source program under the MIT license with source code and training, validation, and testing datasets available at \href{https://github.com/marc-harary/kirigami}{\texttt{https://github.com/marc-harary/kirigami}}.

\begin{figure}[h]
\centering
\includegraphics[angle=-90, width=\textwidth]{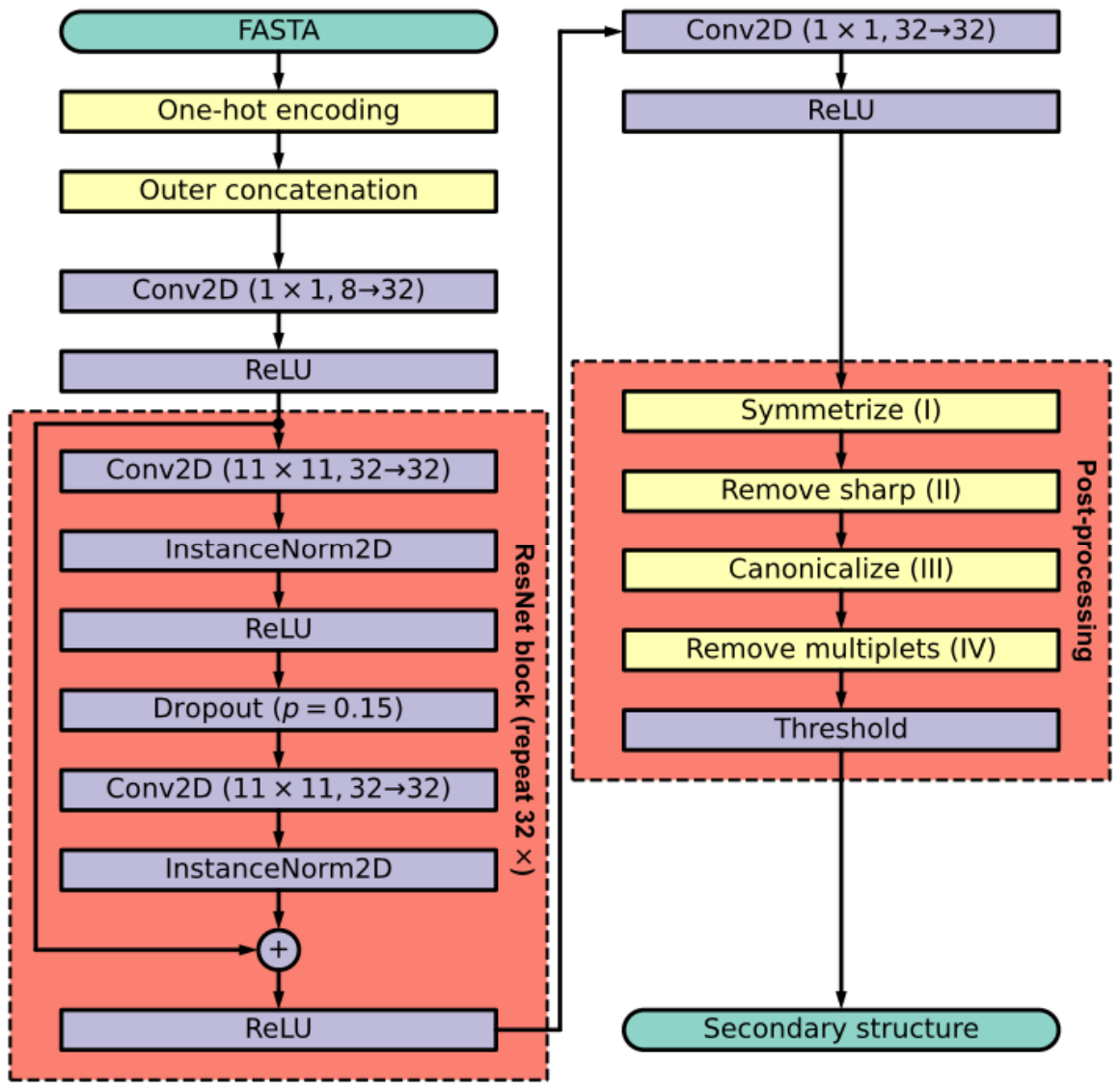}
\caption{Flowchart of the Kirigami pipeline. Green nodes represent input-output values, yellow miscellaneous operations, and blue trainable neural network layers (or closely associated activation functions). Dashed red boxes indicate network submodules. Constraints (I-IV) are indicated in their corresponding nodes.}
\end{figure}

\section{Results}
\subsection{Datasets}

Kirigami was trained and tested on the datasets collected by the authors of SPOT-RNA \cite{4} from the bpRNA database \cite{30}, presently the largest collection of RNA secondary structure annotations. From a range of families that include, among other categories, tRNA, tmRNA, signal recognition particle RNA, and RNase P RNA molecules, 13,419 samples were collected after filtration by CD-HIT-EST \cite{31} for $<80\%$ sequence identity to ensure non-redundancy. The molecules were further divided into sets of 10,814, 1,300 and 1,305 for training, validation, and testing and denoted TR0, VL0, and TS0, respectively. We note that samples from the same RNA families were included in both TR0 and TS0, unlike in some datasets like bpRNA-new used by UFold \cite{18} and MXfold2 \cite{11}.

It should also be noted that higher resolution structures like those in the Protein Databank (PDB) \cite{32} are indeed available and have been used for benchmarking past models like SPOT-RNA, for which a dataset of approximately 200 molecules was collated. However, we found that the comparative scarcity of these data rendered obtaining high model consistency—and hence performing reliable fine-tuning—prohibitively difficult. Model-wise differences in performance were likewise unlikely to yield statistically significant results, prompting our use of the bpRNA database for more robust and reproducible experiments.

\subsection{Overview of pipeline}

Similar to previous studies \cite{4,18}, we interpret RNA secondary structure prediction as an image segmentation task in which 3D input tensors are hierarchically filtered and mapped to outputs of equal height and width. Analogous to pixels in a multichannel raster image, each element in the input corresponds to a particular pair of nucleotides, while the output base-pairing probabilities are analogous to a single-channel grayscale picture (Figure 2A-C). Correspondingly, the backbone of the Kirigami pipeline consists of a 32-block fully convolutional network (FCN) with residual skip connections and, in our case, no pooling layers or dilated kernels. Our network is unique in the size of its convolutional filters; kernel sizes were set to 11 rather than the more standard values of 3 or 5 \cite{4,17,18,22,24,27}. A secondary post-processing module imposes several constraints on the matrix output by the network to remove pairs that (I) are non-symmetric, (II) are non-canonical, (III) create sharp angles, or (IV) form multiplets (see Materials and Methods). 

\begin{figure}[h]
\centering
\includegraphics[angle=-90, width=\textwidth]{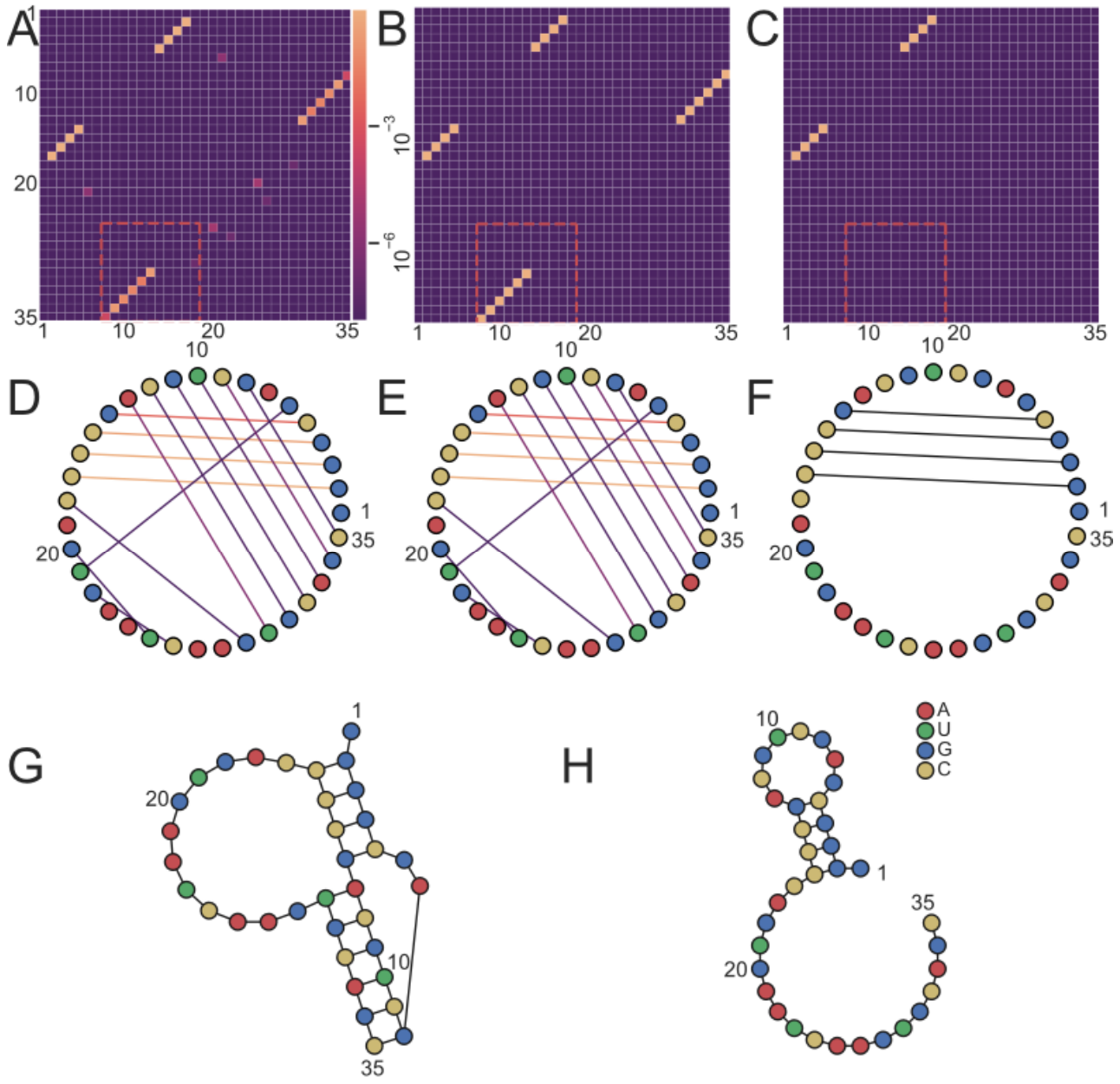}
	\caption{Stages for post-processing using bpRNA\_RFAM\_35590 as an example. A. Constraint I: symmetric predicted base-pairing probabilities $\left( \mathbf{\hat y}_1 \right)$ in log scale. The receptive field of an  kernel is indicated in the red dashed box. B. Adjacency matrix of the ground truth $\left( \mathbf y \right)$. C. Final predicted adjacency matrix $\left(\mathbf{\hat y}_3 \right)$. D. Constraints II: Removal of sharp angles $\left( \mathbf{\hat y}_2 \right)$. In this case none are present. E. Constraint III: Removal of non-canonical pairs $\left( \mathbf{\hat y}_3 \right)$. In this case none are present. F. Constraint IV: Removal of multiplets $\left( \mathbf{\hat y}_4 \right)$—in this case none are present—and thresholding $\left( \mathbf{\hat y}_5 \right)$. G. Ball-and-stick representation of the ground truth $\left( \mathbf y \right)$. Positions assigned using Pseudoviewer \cite{50}. H. Ball-and-stick representation of the final, thresholded structure $\left( \mathbf{\hat y}_5 \right)$.}
\end{figure}

\subsection{Comparison with existing methods}

On TS0, Kirigami achieves an average MCC of 0.706, which is 11.4\%, 17.0\%, 464.4\%, 26.6\%, 41.6\%, 36.6\%, 45.1\%, 34.8\%, and 40.7\% higher than the MCCs of UFold \cite{18}, SPOT-RNA \cite{4}, E2Efold \cite{25}, MXfold2 \cite{11}, RNAfold \cite{7}, PETfold \cite{14}, PPfold \cite{10}, IPknot \cite{17}, and ContextFold \cite{12}, respectively (Table 1, Figure 3A). Improvement is for the most part more pronounced on molecules with pseudoknots, where the average MCC is 9.6\%, 24.7\%, 528.2\%, 38.7\%, 63.4\%, 52.2\%, 62.6\%, 61.4\%, and 37.9\% higher, respectively. All differences are statistically significant $\bigr($Table 1,  $p < 1.76 \times 10^{-11}\bigl)$. 


\begin{table}[h]
\centering
\caption{Performance Comparison of RNA Folding Algorithms}
\begin{subtable}{\textwidth}
\centering
\tiny 
\begin{tabularx}{\textwidth}{lX X X X X}
\toprule
\textbf{Dataset} & \textbf{Kirigami} & \textbf{UFold} & \textbf{SPOT-RNA} & \textbf{E2EFold} & \textbf{MX2fold} \\
\midrule
\textbf{Fully nested} (N=1,176)  & \textbf{0.710} & 0.636 & 0.611 & 0.127 & 0.566 \\
\textbf{Pseudoknotted} (N=129) & \textbf{0.669} & 0.611 & 0.537 & 0.107 & 0.482 \\
\textbf{All} (N=1,305) & \textbf{0.706} & 0.634 (1.76E-11) & 0.604 (5.40E-22) & 0.125 ($\sim$ 0) & 0.558 (4.62E-40) \\
\bottomrule
\end{tabularx}
\end{subtable}

\vspace{1em} 

\begin{subtable}{\textwidth}
\centering
\tiny 
\begin{tabularx}{\textwidth}{lX X X X X}
\toprule
\textbf{Dataset} & \textbf{RNAfold} & \textbf{PETfold} & \textbf{PPfold} & \textbf{IPknot} & \textbf{ContextFold} \\
\midrule
\textbf{Fully nested} (N=1,176)  & 0.509 (2.90E-75) & 0.526 (4.32E-65) & 0.495 (4.38E-83) & 0.536 (1.54E-61) & 0.504 (1.65E-70) \\
\textbf{Pseudoknotted} (N=129) & 0.410 (2.90E-75) & 0.440 (4.32E-65) & 0.412 (4.38E-83) & 0.415 (1.54E-61) & 0.485 (1.65E-70) \\
\textbf{All} (N=1,305) & 0.499 (2.90E-75) & 0.517  (4.32E-65) & 0.487 (4.38E-83) & 0.524  (1.54E-61) & 0.502 (1.65E-70) \\
\bottomrule
\end{tabularx}
\end{subtable}

\end{table}

When considering pseudoknots in isolation from nested pairs, Kirigami obtains a mean MCC of 0.615, which is 58.8\%, 116.6\%, 384.3\%, 213.0\%, 268.3\%, 251.1\%, 401.7\%, 309.4\% and 544.8\% higher than its competitors (Table 3, Figure 3B). The most dramatic difference lay in median MCC; Kirigami obtained a score of 0.797, whereas each of its alternatives exhibited a median less than or equal to 0.

\begin{table}[h]
\centering
\caption{Summary statistics for MCC on pseudoknot pairs in TS0 by model.}
\begin{subtable}{\textwidth}
\centering
\tiny 
\begin{tabularx}{\textwidth}{lX X X X X}
\toprule
\textbf{Statistic} & \textbf{Kirigami} & \textbf{UFold} & \textbf{SPOT-RNA} & \textbf{E2Efold} & \textbf{MXfold2} \\
\midrule
\textbf{Mean} & \textbf{0.615} & 0.387 & 0.284 & 0.127 & 0.196 \\
\textbf{Median} & \textbf{0.797} & 0.00 & 0.00 & -1.13E-3 & -1.41E-3 \\
\bottomrule
\end{tabularx}
\end{subtable}

\vspace{1em} 

\begin{subtable}{\textwidth}
\centering
\tiny 
\begin{tabularx}{\textwidth}{lX X X X X}
\toprule
\textbf{Statistic} & \textbf{RNAfold} & \textbf{PETfold} & \textbf{PPfold} & \textbf{IPknot} & \textbf{ContextFold} \\
\midrule
\textbf{Mean} & 0.167 & 0.175 & 0.122 & 0.150 & 0.095 \\
\textbf{Median} & -1.87E-3 & -1.73E-3 & 0.00 & 0.00 & -1.25E-3 \\
\bottomrule
\end{tabularx}
\end{subtable}

\end{table}

\subsection{Ablation studies}
To quantify the effect of kernel size on performance, we retrained otherwise identical networks with varying kernel sizes. In other words, fixing  $N_{\text{channels}} = N_{\text{blocks}} = 32$ (see Materials and Methods) and the training procedure described below, networks were trained with smaller kernel sizes of $k= 9$, $k=7$, and  $k=5$. Respectively, these resulted in MCCs of 0.680 $\left( p = 1.35 \times 10^{-2} \right)$, 0.681 $\left( p = 1.62 \times 10^{-2} \right)$, and 0.654 $\left( 9.87 \times 10^{-7} \right)$, equivalent to percent decreases of 3.8\%, 3.6\%, and 7.4\% with respect to the default configuration (0.706, Table 2). One confounding factor, however, is network size; larger kernels entail a larger number of total parameters and in turn higher model capacity. We therefore tested a network with  $N_{\text{channels}} = 32$,  $N_{\text{blocks}}= 64$, and $k = 5$, which had approximately 3.3 million parameters in comparison to the 3.2 million parameters of the network with hyperparameters corresponding to $N_{\text{channels}} = 32$, $N_{\text{blocks}} = 32$ and $k = 7$. This modified configuration exhibited a 5.42\% decrease in MCC relative to the default configuration from 0.706 to 0.668. Hence, the data consistently indicate that kernel size—rather than total parameter number \textit{per se}—improve prediction accuracy.

\begin{table}[h]
\centering
\caption{Mean MCC on TS0 for Kirigami by kernel size and number of blocks.}
\tiny
\begin{tabularx}{\textwidth}{lX X X X X}
\toprule
$\mathbf{ \left(k, N_{\text{blocks}}\right)}$ & \textbf{(11, 32)} & \textbf{(9, 32)} & \textbf{(7, 32)} & \textbf{(5, 32)} & \textbf{(5, 64)} \\
\midrule
\textbf{Fully nested} (N=1,176) & \textbf{0.710} & 0.687 & 0.687 & 0.663 & 0.674 \\
\textbf{Pseudoknotted} (N=129) & \textbf{0.669} & 0.617 & 0.625 & 0.572 & 0.616 \\
\textbf{All} (N=1,305) & \textbf{0.706} & 0.680 & 0.681 (1.35E-2) & 0.654 (9.18E-7) & 0.668 (3.06E-4) \\
\bottomrule
\end{tabularx}
\end{table}

\subsection{Biophysical analysis}
Using a software tool available in the ViennaRNA \cite{7} suite, we computed the average Gibbs free energy per mole per base (kJ/mol/b) as approximated by NNDB \cite{13} of both the ground truth secondary structures and those predicted by each SOTA model, then computed the difference $\Delta G$ between the two estimates. Because multiplets or sharp angles either cause the program to crash or return a meaninglessly high output, they were removed (only for biophysical analysis) from both ground truth structures and those predicted by models that do not natively enforce such constraints. E2Efold \cite{25} on average massively underestimates ground truth stability, whereas the inverse holds for MXfold2 \cite{11}, RNAfold \cite{7}, PETfold \cite{14}, IPknot \cite{17}, and ContextFold \cite{12}. Only the neural models Kirigami, UFold \cite{18}, and SPOT-RNA \cite{4}, along with the maximum expected accuracy (MEA) model PPfold \cite{10}, exhibit medians approximately equal to the ground truth (Figure 4A).

We then regressed the model-wise MCC onto $\Delta G$ (Figure 4B). Expectedly, the correlations were highest for the MFE models, namely RNAfold $\left( r= 0.528, \ p=1.11 \times 10^{-94} \right)$, PETfold $\left( r=0.459, p=5.35 \times 10^{-69} \right)$, and MXfold2 $\left( r=0.583, \ p=1.44 \times 10^{-119}\right)$, with ContextFold $\left(r=0.313, \ p=5.61 \times 10^{-31}\right)$ being an exception. Lower (albeit still statistically significant) correlations were observed for Kirigami $\left( r=0.251, \ p=3.84 \times 10^{-20}\right)$, UFold $\left( r=0.293, \  p=2.91 \times 10^{-27} \right)$, and SPOT-RNA $\left( r= 0.270, \ p=2.78 \times 10^{-23} \right)$. As to the MEA algorithms, PPfold exhibited the lowest correlation $\left( r= 0.118, \ p=1.91 \times 10^{-5} \right)$ while IPknot’s correlation $\left( r= 0.292, \ p=5.58 \times 10 ^{-27} \right)$ was expectedly closer to those of the neural models than to its biophysical counterparts.

\begin{figure}[h]
\centering
\includegraphics[angle=-90, width=\textwidth]{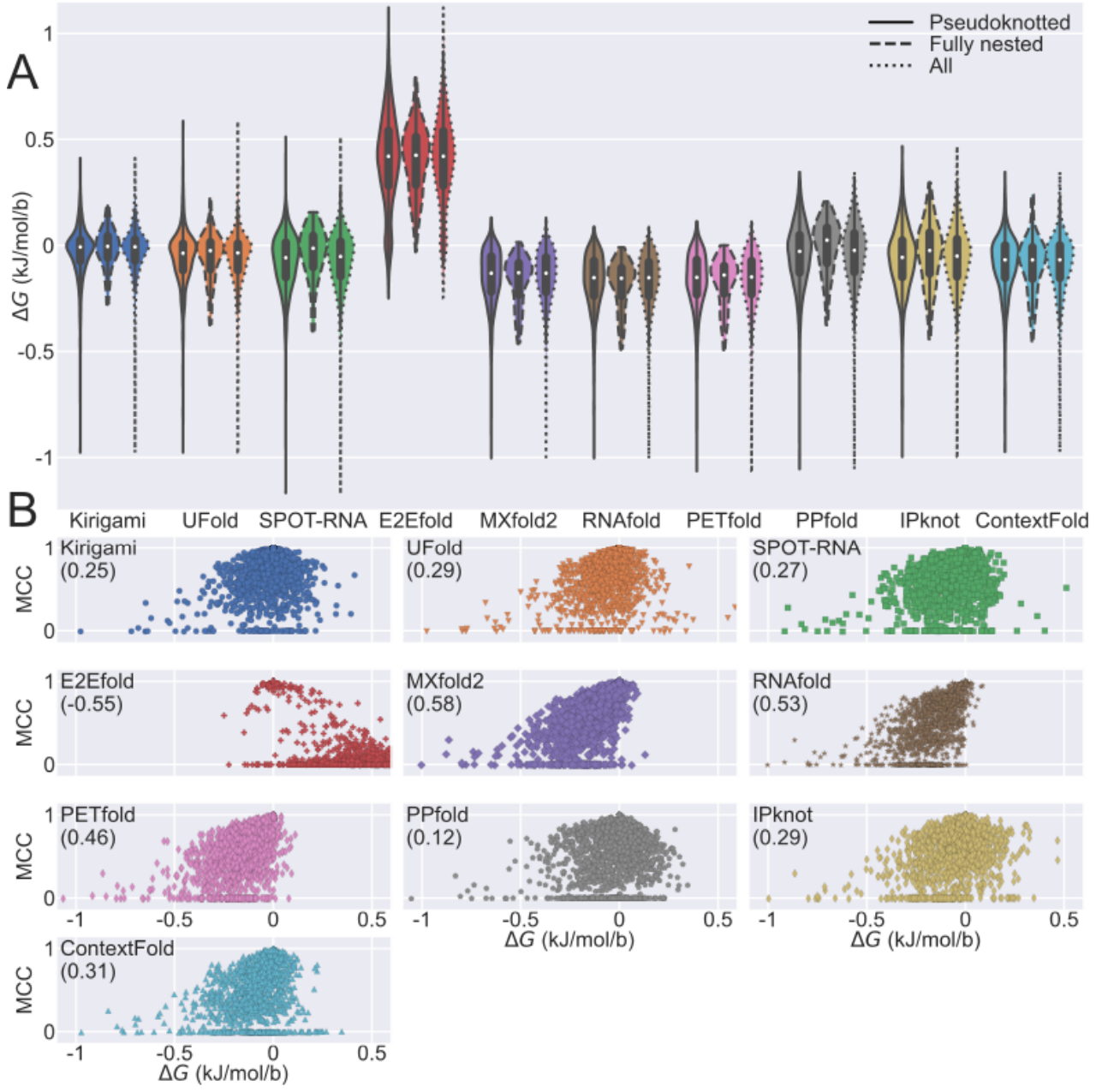}
\caption{A. Difference between the Gibbs free energy ($\Delta G$) of the predicted structure and that of the ground truth by model as predicted by ViennaRNA. B. MCC versus $\Delta G$ by model with the corresponding correlation coefficient.}
\end{figure}

\section{Discussion and Conclusion}
In this work, we develop Kirigami, a deep learning model for RNA secondary structure prediction. On a large benchmark dataset of 1,305 RNA structures from the bpRNA \cite{30} library, it achieves a mean MCC of 0.706, which is 11.4\%, 26.6\%, and 17.0\% higher than the SOTA methods UFold \cite{18}, MXfold2 \cite{11}, and SPOT-RNA \cite{4}, respectively, with an MCC of 0.615 on pseudoknots that is 58.8\%, 213.0\%, and 116.6\% higher, respectively.

Along with these three neural networks, Kirigami significantly outperforms other secondary structure prediction algorithms and particularly MFE models, the growing superiority of neural to biophysical models paralleling the “neuralization” of many aspects of protein structure prediction \cite{33}. The causes of the discrepancy are unclear; given that neural networks act as universal function approximators \cite{34,35}, an outstanding question in secondary structure prediction is the nature of the function estimated when models are trained on large datasets. One possibility is that neural networks learn the energy landscape of RNA structures and are simply doing so to a greater degree of accuracy than is possible via the relatively small number of parameters that inform most biophysical models \cite{7, 8, 9, 12, 13}. As is shown in Figure 3A, UFold, SPOT-RNA, and Kirigami predict structures that not only exhibit the highest MCC score, but also the smallest values of $\Delta G$.

However, such improvements cannot be explained by neural models simply being more capable of optimizing the landscape more efficiently, given that most MFE models still typically overestimate molecular stability. Nor can it be the case that Kirigami and other neural algorithms outperform biophysical methods by avoiding an inductive bias towards strictly energetically minimizing solutions—which, as noted above, can often limit MFE algorithms \cite{14}—given that correlations between the accuracy of the thermodynamic stability of the networks' predictions and their MCCs are lower. Thermodynamic accuracy, in other words, cannot in general mediate between model type and performance on secondary structure prediction. The nature of the mappings learned by Kirigami and other neural models may therefore remain a topic of interest for future inquiry.

As to the optimal architecture for neural networks in this domain, we hypothesize that large-kernel fully convolutional networks \cite{27} are ideal for the one of the most challenging aspects of RNA secondary structure prediction, namely the unique topology of output labels. Due to base stacking, most molecules consist of a small quantity of large stems separated by unstructured loops \cite{30,32}, and due also to the rarity of multiplets, most hydrogen bonds will be isomorphic to long, linear neighborhoods of contiguous positive labels that are one pixel in width when the structure is embedded as an image-like tensor \cite{4,18}. While large feature patches are not uncommon in many computer vision tasks, image segmentation does not typically involve the detection of wider, two-dimensional connected components characterized by such high spatial frequency and fine detail. Unusually large convolutional kernels like those in Kirigami may thus be more effective at delineating stems from nearby unstructured regions of molecules than kernels with commoner \cite{4,17,18,22,24,27}, smaller receptive fields that can only detect more localized structure. For example, in Figure 2A-C, molecule bpRNA\_RFAM\_35590 has a stem consisting of 6 bonds that, along with up to five bases of the surrounding hairpin loop, can fit well within the boundaries of an 11-pixel kernel. Kirigami is hence able to detect both the location and extent of the loop with high accuracy, exhibiting low activations for the edges surrounding the stem (Figure 2A). Similarly, pooling layers for down-sampling, in spite of their utility for models like U-Net \cite{26} designed for more standard image segmentation tasks, may be less adept at precisely locating the narrow strips of positive labels that are characteristic of embedded RNA secondary structures.

Despite the success of Kirigami’s simple ResNet architecture and the recent advances in convolutional neural networks for computer vision, transformer-based models \cite{36} have proliferated in many other fields of computational biology \cite{37}. Future directions for research into neural approaches for RNA structure may ultimately include attention mechanisms, which have proven highly performant in protein structure prediction tasks \cite{37}.

Still, the most challenging aspect of RNA computational biology is the paucity of high-quality data \cite{4}. As shown in the recent CASP15 challenge, deep learning approaches are not yet up to the accuracy of statistical or empirical methods for many structure prediction tasks. Recent improvements in RNA experimental structure determination techniques such as cryogenic electron microscopy \cite{32} will eventually expand existing databases and enable the development of more sophisticated machine learning models.

\section{Materials and methods}
\subsection{Input features and output labels}

More formally, we consider the structure of a molecule to be a weighted graph in which edges represent base pairing probabilities (Figure 2D-F) and nodes, embedded as feature vectors, represent nucleotide residues. Secondary structure prediction thus amounts to inferring the corresponding adjacency matrix, with intermediate, ``hidden'' representations of the molecule being updated via permutation-invariant layers in the form of two-dimensional convolutions. The topologies of the network and the tensors that it manipulates, however, remain identical to those found in standard computer vision models.

As input features, each $i$th node $\mathbf h_i$ in the molecular graph consists of the nucleic acids adenine (A), cytosine (C), guanine (G), and uracil (U) represented by the standard basis vectors $\mathbf e_1$, $\mathbf e_2$, $\mathbf e_3$, and $\mathbf e_4$, respectively, while degenerate bases \cite{38} are mapped to the corresponding averages thereof $\bigl($e.g., ketones are represented by $\frac{1}{2} \left( \mathbf e_1 + \mathbf e_4 \right) \bigr)$. The relationship between each pair of nucleotides $\left( i, j\right)$, in order to be iteratively refined by the network into a corresponding base pairing probability $\hat{y}_{ij}$, is initially represented by simply concatenating $\mathbf h_i$ and $\mathbf h_j$ to a form a vector $\mathbf x_{ij} \in \mathbb R^8$. Therefore, the final input feature for all $L^2$ ordered pairs of nucleotides is a tensor $\mathbf x \in \mathbb R^{8 \times L \times L}$, analogous to a square eight-channel image.

As an output label, the RNA secondary structure can be straightforwardly embedded as a binary adjacency matrix $\mathbf y \in \left\{ 0, 1 \right\}^{L \times L}$, where $y_{ij} = 1$ if and only if the $i$th and $j$th nucleotides form a base pair.

\subsection{Network architecture}
The Kirigami pipeline consists of a 32-block residual \cite{22} fully convolutional network (FCN) \cite{27}, meaning that it is does not contain multilayer perceptron \cite{23} or recurrent neural network \cite{20,21} layers. The principal hyperparameters of the network consist of the triple $\left( N_{\text{channels}}, \ N_{\text{blocks}}, \ k \right)$, where $N_{\text{channels}}$ denotes the number of intermediate input channels in the network, $N_{\text{blocks}}$ the number of ResNet blocks \cite{22}, and $k$ the size of the kernels of the convolutions in each block. Irrespective of the hyperparameters, the first two layers consist of a convolution by a $1 \times 1$ kernel followed by ReLU activation, which expand the input tensor $\mathbf x \in \mathbb R^{8 \times L \times L}$ to an intermediate feature map $\mathbf z^{(1)} \in \mathbb R^{N_{\text{channels}} \times L \times L}$. Each message-passing block resembles that of a typical ResNet insofar as each is comprised of two 2D convolutions—both with $k \times k$ kernels and followed by ReLU activation and 2D instance normalization \cite{39}—with a single dropout layer separating these two halves. Denoting the output of the $i$th block $\mathbf z^{(i+1)} \in \mathbb R^{N_{\text{channels}} \times L \times L}$, a residual skip connection adds the initial input to the block $\mathbf z^{(i)} \in \mathbb R^{N_{\text{channels}} \times L \times L}$ back to the output of the second normalization layer (Figure 1). The final layers of the network consist of a $1 \times 1$ convolution and sigmoid activation to reduce the dimension and range of the output to the final label $\mathbf{\hat y}_0 \in \left[ 0, 1 \right]^{L \times L}$. All convolutions have strides and dilations of 1 \cite{24}.

Grid search, as constrained by hardware limitations, revealed that optimal performance was obtained by setting $N_{\text{channels}}$ to 32, $N_{\text{blocks}}$ to 32, and $k$ to 11.

\subsection{Post-processing}
RNA contact maps \cite{40,41} bear a close resemblance to, but are ultimately distinct from, secondary structure adjacency matrices. Whereas in the former, any pair of nucleotides can be considered in contact so long as their interatomic distance is within 8Å, two nucleotides are considered paired in the latter if and only if they are paired by specific hydrogen bonding interactions. Hence, predicted RNA secondary structures must obey certain constraints \cite{4}, namely that, if nucleotides $i$ and $j$ are paired, they are assumed to exhibit:

\begin{enumerate}
	\item[(I)] Symmetry: $j$ is also paired to $i$;
	\item[(II)] Minimal loop length: $\left| i - j \right| \geq 4$;
	\item[(III)] Watson-Crick or Wobble pairing: the nucleotide pair must be A:U, C:G, or G:U;
	\item[(IV)] Absence of multiplets: neither  nor  can be paired with any other base.
\end{enumerate}

These collectively necessitate constraint-satisfaction \cite{42} via post-processing. To enforce Constraint (I), the output matrix $\mathbf{\hat y}_0$ is simply averaged with its own transpose, yielding a new matrix $\mathbf{\hat y}_1 := \frac{1}{2} \left( \mathbf{\hat y}_0 + \mathbf{\hat y}_0^\intercal \right)$ (Figure 2A), while satisfying Constraint (II) is straightforwardly accomplished by zeroing out the entries of the upper and lower diagonals to produce another intermediate matrix  $\mathbf{\hat y}_2$ (Figure 2D displays $\mathbf{\hat y}_2$ as a graph).

Canonicalization was enforced by encoding the bases A, C, G, and U as the prime numbers 2, 3, 5, and 7, respectively. Prime number-encoding converts the RNA sequence into a vector $\mathbf v \in \left\{ 2, 3, 5, 7 \right\}^L$ such that the matrix $\mathbf M := \mathbf v \otimes \mathbf v$ uniquely represents each base pair permutation, with 14, 15, and 35 corresponding to A:U, C:G, and G:U pairs, respectively. Enforcing (III) can thus be performed by applying the differentiable entry-wise matrix multiplication $\mathbf{\hat y}_3 := \mathbf{\hat y}_2 \odot \mathds 1_{\left\{14, 15, 25 \right\}} \left( \mathbf M \right)$, where $\mathds{1}_{\left\{14, 15, 25\right\}}$ denotes the indicator function associated with the three canonical pairs in the prime number-encoding (Figure 2E).

Somewhat similar to the post-processing procedure used by SPOT-RNA \cite{4}, Constraint (IV) is then enforced by a greedy algorithm that proceeds as follows. All pairs are sorted in descending order by score in $\mathbf{\hat y}_3$ and then iterated over. For each pair $(i, j)$, if there exists a pair $(i, k)$ or $(j, k)$ already in the solution set $A$, $(i, j)$ is discarded. After the conclusion of the loop, a new symmetric matrix $\mathbf{\hat y}_4$ is initialized such that only the pairs of indices in $A$ are set to 1. Notably, in contrast to dynamic programming \cite{15}, pseudoknots are able to be returned.

As in both SPOT-RNA and UFold \cite{18}, insignificant pairs are excluded by applying a threshold $T$ computed via grid search on VL0 to produce the final output $\mathbf{\hat y}_5 := \mathds{1}_{y \geq T} \left( \mathbf{\hat y}_4 \right)$ (Figure 2CF). The final result was a cutoff of $T = 0.455$.

\subsection{Training procedure}
The network was implemented in the PyTorch library \cite{43} with the overall software package written according to the Lightning specification. Separate loops for training, validation, and testing were thus automatically delineated as separate class methods, with the validation loop called at the conclusion of each fifth iteration of the training loop.

As to the latter, training of Kirigami was conducted for 100 epochs over the 10,814 molecules in TR0 using the Adam gradient descent algorithm \cite{44} with a learning rate of $1 \times 10^{-4}$ for binary cross-entropy loss minimization and a dropout rate \cite{45} of 0.15. Due to the hardware consisting of a compute node with a single NVIDIA GeForce RTX 3090, batch size was limited to 1; gradient accumulation \cite{46} was therefore conducted across every 64 batches. Of Constraints (I-IV) described above, only (I) was enforced during training; no thresholding was performed or metrics evaluated aside from cross-entropy loss.

Validation epochs consisted of iterating over VL0, enforcing Constraints (I-IV), thresholding the output matrix across a grid of 100 evenly spaced values along the interval $\left[ 0, 1 \right]$ (Figure 2), and evaluating the classification metrics at each threshold. For each sample in VL0, the Matthews Correlation Coefficient (MCC) \cite{47} along with other statistics was computed by the formula below at successive values of $T$, then cached in a table initialized at the beginning of each validation epoch. Following the call to the validation loop method, the maximizing argument of $T$ with respect to the mean MCC across VL0 was computed via a final round of grid search and written to the network as a parameter. Network weights were checkpointed if the maximum mean MCC exceeded that of previous epochs.

The training procedure can be reproduced via scripts in the corresponding GitHub repository (see above). More straightforwardly, the test loop, which replicates the results below (Table 1), can be called using pretrained weights made publicly available in the same codebase. In order to facilitate the use of Kirigami on molecules outside of TS0, a prediction loop for reading in FASTA files and outputting corresponding dot-bracket files was also developed, the conversion between the output matrix and the dot-bracket string computed via an application of Nussinov’s recurrence relation for identifying pseudoknots as described by \cite{49}.

\subsection{Evaluation metrics}
Performance for each model was measured in terms of Matthews correlation coefficient (MCC) as follows. Representing the ground truth and predicted secondary structures of a molecule of length $L$, the upper triangles of the symmetric binary matrices $\mathbf y, \mathbf{\hat y} \in \left\{ 0, 1 \right\}^{L \times L}$ were flattened into feature vectors $\mathbf v, \mathbf{\hat v} \in \left\{ 0, 1 \right\}^{\frac{(L-1)L}{2}}$, respectively. Critically, the upper triangles were offset by one element to exclude the diagonal from the computation. The MCC between $\mathbf v$ and $\mathbf{\hat v}$ was then computed via the formula
\begin{equation}
	MCC = \frac{TN \times TP - FN \times FP}{\sqrt{(TP + FP)(TP + FN)(TN+FP)(TN+FN)}},
\end{equation}
where $TP$ denotes the number of true positives, $FP$ false positives, $TN$ true negatives, and $FN$ false negatives \cite{47}. To evaluate the MCC on pseudoknots specifically, the indices $i_1, i_2, \ldots, i_n$ of  $\mathbf v$ that corresponded to non-nested pairs were used to create new vectors $\mathbf v_p := \left(v_{i_1}, v_{i_2}, \ldots, v_{i_n}  \right)$ and $\mathbf{\hat v}_p := \left( \hat v_{i_1}, \hat v_{i_2}, \ldots, \hat v_{i_n} \right)$, between which the MCC was then computed according to the formula above. The percentage increase $\Delta$ in MCC of Kirigami over other models was simply computed via
\begin{equation}
	\Delta = 100 \% \times \frac{MCC_{\text{Kirigami}} - MCC_{\text{model}}}{MCC_{\text{model}}}.
\end{equation}

\section*{Acknowledgements}

We acknowledge the invaluable support, guidance, and insight of Anna Marie Pyle at Yale University in formulating our project.

\bibliographystyle{plain}
\bibliography{refs}

\end{document}